\documentclass[preprint]{aastex}
\usepackage{psfig}

\begin{document}
\parskip=0pt

\let\deg=\arcdeg

\def\ie{{\it i.e.,\ }}
\def\eg{{\it e.g.,\ }}
\def\qv{{\it q.v.,\ }}
\def\cf{{\it cf.\ }}
\def\etal{{\it et al.}}
\def\gtrapprox{\;\lower 0.5ex\hbox{$\buildrel >
    \over \sim\ $}}             
\def\lessapprox{\;\lower 0.5ex\hbox{$\buildrel < \over \sim\ $}}
\def\msol{\ifmmode {\>M_\odot}\else {$M_\odot$}\fi}
\def\Omegab{\ifmmode {\Omega_{\rm baryon}}\else {$\Omega_{\rm baryon}$}\fi}
\def\pyr{\ifmmode {\>{\rm\ yr}^{-1}}\else {yr$^{-1}$}\fi}
\def\psec{\ifmmode {\>{\rm\ s}^{-1}}\else {s$^{-1}$}\fi}
\def\kms{\ifmmode {\>{\rm km\ s}^{-1}}\else {km s$^{-1}$}\fi}
\def\psqcm{\ifmmode {\>{\rm cm}^{-2}}\else {cm$^{-2}$}\fi}
\def\pcubcm{\ifmmode {\>{\rm cm}^{-3}}\else {cm$^{-3}$}\fi}
\def\phoflux{\ifmmode{{\rm phot\ cm}^{-2}{\rm\ s}^{-1}}\else {phot
cm$^{-2}$ s$^{-1}$}\fi}

\def\dex#1{10$^{#1}$}
\def\tdex#1{$\times$10$^{#1}$}
\def\cmm#1{\,cm$^{-#1}$}
\def\vlsr{\ifmmode{\>{v_{\rm LSR}}}\else{v$_{\rm LSR}$}\fi}
\def\vgsr{v$_{\rm GSR}$}
\def\TB{T$_{\rm B}$}
\def\HI{\protect\ion{H}{1}}
\def\nh{\ifmmode{n_{\rm H}}\else{$n_{\rm H}$}\fi}
\def\nhi{\ifmmode{n_{\rm HI}}\else{$n_{\rm HI}$}\fi}
\def\Nh{N$_{H}$}
\def\Mgas{\ifmmode{M_{gas}}\else{$M_{gas}$}\fi}
\def\Mdm{\ifmmode{M_{dm}}\else{$M_{dm}$}\fi}
\def\Mhi{\ifmmode{M_{\rm HI}}\else{$M_{\rm HI}$}\fi}
\def\Nhi{\ifmmode{N_{\rm HI}}\else{$N_{\rm HI}$}\fi}
\def\rh{r$_{H}$}
\def\rhi{r$_{\rm HI}$}
\def\be{\begin{equation}}
\def\ee{\end{equation}}
\def\bea{\begin{eqnarray}}
\def\eea{\end{eqnarray}}

\title{Are Compact High-Velocity Clouds Extragalactic Objects?}

\author{Philip R. Maloney\altaffilmark{1} \& Mary
  E. Putman\altaffilmark{2,3}}  
\affil{Center for Astrophysics and Space Astronomy, University of Colorado\\  
       Boulder, CO 80309-0389}

\altaffiltext{1}{maloney@casa.colorado.edu}
\altaffiltext{2}{Hubble Fellow}
\altaffiltext{3}{mputman@casa.colorado.edu}

\begin{abstract}
Compact high-velocity clouds (CHVCs) are the most distant of the HVCs
in the Local Group model and, at $d\sim$ 1 Mpc, they have HI volume
densities of $\sim 3 \times 10^{-4}$ \pcubcm. Clouds with these volume
densities and the observed column densities $\Nhi \sim 10^{19}$
\psqcm\ will be largely ionized, even if exposed only to the
extragalactic ionizing radiation field. Here we examine the
implications of this process for models of CHVCs. We have modeled the
ionization structure of spherical clouds (with and without dark matter
halos) for a large range of densities and sizes, appropriate to CHVCs
over the range of suggested distances, exposed to an extragalactic
ionizing photon flux $\phi_i\sim 10^{4}$ \phoflux. Constant-density
cloud models in which the CHVCs are at Local Group distances have
total (ionized plus neutral) gas masses $\sim 20-30$ times larger than
the neutral gas masses, implying that the gas mass alone of the
observed population of CHVCs is $\sim 4 \times 10^{10}$ \msol. With a
realistic (10:1) dark matter to gas mass ratio, the total mass in such
CHVCs is a significant fraction of the dynamical mass of the Local
Group, and their line widths would greatly exceed the observed $\Delta
V$. Self-consistent models of gas in dark matter halos fare even more
poorly; they must lie within approximately 200 kpc of the Galaxy, and
(for a given distance) are much more massive than the corresponding
uniform density models. We also show that exponential neutral hydrogen
column density profiles are a natural consequence of an external
source of ionizing photons, and argue that these profiles cannot be
used to derive model-independent distances to the CHVCs. These results
argue strongly that the CHVCs are not cosmological objects, and are
instead associated with the Galactic halo.
\end{abstract}


\section{Introduction}
The anomalous velocity clouds of neutral hydrogen known as the
high-velocity clouds (HVCs) may represent the continuing infall of
matter onto the Local Group (\eg Oort 1966; Verschuur 1969; Blitz
\etal\ 1999). The distances to the majority of these clouds remains
unknown, but in the Local Group scenario the distances would range
from a few kpc (for gas currently accreting onto the Galaxy) to beyond
Andromeda ($\sim 1$ Mpc). HVCs range in size from $\sim 0.1 - 100$
deg$^{2}$, and it is possible that the sizes reflect their distance
from the Galaxy. In particular, some of the HVCs are both compact and
isolated from extended emission, and show many similar HI properties
to dwarf galaxies. These are referred to as compact HVCs (CHVCs) and
may be the pristine building blocks of the Local Group at $\sim 1$
Mpc.

The survival and composition of these CHVCs against photoionization
depends on their volume densities (and therefore distances) and the
strength of the extragalactic ionizing photon flux. The presence of an
extragalactic ionizing radiation field is inferred from several sets
of observations. Maloney (1993) used a deep \HI\ observation of NGC
3198 (van Gorkom 1993), which shows a sharply truncated neutral
hydrogen disk at $\Nhi \approx 5 \times 10^{19}$ \psqcm, to infer that
the flux in the energy range $hv = 13.6$ to $\sim 200$ eV is
approximately $\phi_i = 10^4$ \phoflux. Estimates from the proximity
effect at low redshift (Kulkarni \& Fall 1993; Scott \etal\ 2002),
upper limits from sensitive H$\alpha$ observations (Madsen \etal\
2001; Weymann \etal\ 2001), and theoretical estimates based on known
sources (Haardt \& Madau 1996; Shull \etal\ 1999) are in agreement
with a value $\phi_i\sim 0.5-1.5\times 10^4$ \phoflux.

In this paper we examine the ionization state of the compact
high-velocity clouds at various distances, subject to an extragalactic
ionizing radiation field. In the original Local Group model, the CHVCs
are confined by dark matter halos, with $M_{dm}/M_{gas}\approx 10$. In
a more recent variant of this model, it is suggested that the CHVCs
are condensations in an intragroup medium, bound by external
pressure (Blitz, private communication). We therefore consider both
constant gas density models and 
dark matter-dominated models.
The \HI\ properties of the CHVCs are briefly reviewed in section 2 and
the photoionization model is described in section 3. The results of
the models are presented in section 4, and in section 5 we discuss the
implications of the results.

\section{HI Data}
The \HI\ properties of the CHVCs are taken from the \HI\ Parkes
All-Sky Survey (HIPASS) HVC catalog of Putman \etal\ (2002). This is a
catalog of high-velocity clouds with declination $\delta < +2^\circ$
and $|\vlsr|\le 500$ \kms, excluding emission with $|\vlsr|< 90$
\kms\ to avoid confusion with Galactic emission. The spatial
resolution of the survey is $15.5'$ and the velocity resolution is 26
\kms. The $5\sigma$ column density sensitivity of the HVC HIPASS data
is $2\times 10^{18}$ \psqcm, which we adopt as a cutoff. Barnes \etal\
(2001) give a full description of HIPASS. The original CHVC catalog of
Braun \& Burton (1999) was based on data of lower resolution and
sensitivity, and 50\% of the CHVCs in the region of overlap between
the two catalogs were reclassified based on the HIPASS data. In order
for a high-velocity cloud to be considered a compact and isolated CHVC
in the HIPASS catalog, it must have a diameter at 25\% of the peak
column density less than 2\deg,
and must not be elongated in the direction of any extended
emission. The typical parameters of a HIPASS CHVC are an angular size
of 0.36 deg$^{2}$ (to the sensitivity limit), a total \HI\ flux of
19.9 Jy \kms, and a typical peak column density of $\Nhi\sim 10^{19}$
\psqcm. There are a total of 179 CHVCs cataloged from HIPASS and when
this is combined with the northern sky data from the Leiden-Dwingeloo
survey there are approximately 250 CHVCs in total (de Heij, Braun \&
Burton 2002). The linewidth distribution shows a prominent peak at a
FWHM $\Delta V=25\kms$ (de Heij, Braun \& Burton).

For most of the results in this paper, we have used the typical CHVC
parameters quoted above. This is a reasonable assumption, since the
distributions in column density, angular size, and velocity are fairly
sharply peaked. An increase or decrease of less than a factor of three
in column density away from the peak value produces a drop by a factor
of five or more in the number of clouds\footnote{Sensitivity starts to
affect the column density distribution for column densities not too
far below the peak, and so the drop-off with decreasing column below
the peak is not reliably established. However, since clouds with
column neutral hydrogen column densities smaller than the peak value
will be even more highly ionized than clouds with our typical
parameters, the actual number of clouds at the low column density end
is irrelevant for the purposes of this paper.}, and similarly for the
cloud areas (\cf Figures 12 and 13 of Putman \etal\ [2002]). The
linewidth distribution of de Heij \etal\ (2002) is less sharply peaked
than the column density or angular size distributions, and is skewed
to the low-velocity end: $70\%$ of the observed clouds have a FWHM
$\Delta V=25\kms$ {\it or less}; less than $10\%$ have $\Delta V$ in
excess of $42 \kms$. Despite the well-defined typical values, in order
to verify that our conclusions are robust to uncertainties in the
cloud parameters, we have also included models in which the peak
central column density is up to an order of magnitude larger than the
typical CHVC value, and the linewidth is doubled to $\Delta V=50\kms$
(see \S 4.2.2).

\section{Model}
The photoionization models of CHVCs have been calculated with the code
described by Maloney (1993), modified to allow for the spherical
geometry of the CHVCs. (Modest -- tens of percent -- departures from
sphericity will have no significant impact on the model results.)
Spherical geometry raises the ionization rates substantially over
plane-parallel models with the same total hydrogen density;
integration of the ionization rates over angle is done using 10-point
Gaussian quadrature. The photoionization equilibrium of a CHVC exposed
to an isotropic background radiation field is calculated iteratively,
including the effects of the diffuse radiation. Thermal equilibrium at
T $\sim 10^4$ K is assumed, as is appropriate for gas in the warm
neutral phase. An ionizing photon flux $\phi_i\approx 10^4$ \phoflux\
has been assumed; a flux a factor of two smaller would reduce the
total gas masses by a factor of $\approx 1.5$ (see below). A radial
grid with 100 depth points (with logarithmic spacing) was used; we
have verified that this provides adequate accuracy (less than $1\%$
error in the total column density for a fixed neutral column
density). In all cases we have assumed that the only ionizing photon
flux is due to the extragalactic background; depending on the fraction
of ionizing photons that escape from the Milky Way, the Galactic
ionizing radiation field could dominate for CHVC distances
$d\lessapprox 100$ kpc (Maloney \& Bland-Hawthorn 1999).

\section{Results}
As noted in the introduction, we have considered both models with
uniform gas (total hydrogen) density \nh\ and models where the gas is
confined by dark matter potential wells. We first discuss the constant
\nh\ models, as these contain much of the relevant physics, and then
the models that include dark matter.

\subsection{Uniform density models}
If the CHVCs were at local group distances ($\sim 1$ Mpc), their \HI\
volume densities would be $\nhi\sim 10^{-4}\pcubcm$. We therefore
consider total hydrogen densities \nh\ ranging from $10^{-4}$ to
$10^{-2}$ \pcubcm; as we will see, the low-density end of this range
is untenable for reasonable HVC distances. The model results (total
hydrogen columns, gas masses, etc.) are given in Table 1.

In Figure 1, we plot the column density of total hydrogen that is
required to produce a neutral hydrogen column $\Nhi=10^{19}\psqcm$ as
a function of the total hydrogen density, when the model CHVCs are
exposed to our fiducial ionizing photon flux. The solid curve is for our
fiducial ionizing photon flux and the dashed curve is for $\phi_i$
reduced by a factor of two. At the low density end of the range, the
ratio \Nh/\Nhi\ is enormous, $\sim 100-200$, in consequence of the
small neutral fraction of gas at this density even if exposed only to
the extragalactic ionizing background. At the highest densities, this
ratio has declined to $\approx 2.5-3$, as the gas is substantially
neutral over much of the total column.

Since we know both the neutral and total hydrogen column densities as
well as the volume densities, we can calculate the gas masses directly
from the models. However, in order to compare the results to
observations, we need to assign distances to the model clouds. To do
this, we adopt the following procedure. For each model, we determine
the radius at which the projected \HI\ column drops to the HIPASS
sensitivity limit of $\Nhi=2\times 10^{18}\psqcm$. We then set the
distance by requiring that the apparent angular size (using this
radius) matches the typical value found for the CHVCs in the HIPASS
survey, for which the angular radius $\Delta\theta\sim 0.34$
degrees. This gives the result $d_{\rm HVC}=169 R_{18.3}$ kpc, where
$R_{18.3}$ is the model radius (in kpc) at the $\Nhi=2\times
10^{18}\psqcm$ level. 

In Figure 2 we show both the total gas masses \Mgas\ and the apparent
masses \Mhi, the latter derived from the neutral hydrogen column
densities and the apparent (as measured in \HI) sizes, as a function
of the cloud distance and total gas volume density, for our fiducial
ionizing photon flux. As is evident from Table 1, models in which
the total gas density is less than $\nh\sim 3\times 10^{-3}$ are ruled
out, as such clouds are so large that they are required to lie outside
of the Local Group, and the gas masses become absurdly large. At the
maximum acceptable distance ($d\sim 1$ Mpc) the total gas mass is an
order of magnitude larger than the apparent \HI\ mass.

Figure 3 shows the same quantities as in Figure 2, but we now show the
results only for clouds within the Local Group ($d \le 1$ Mpc). In
addition, this figure also shows the gas masses for $\phi_i$ a factor
of two smaller than our assumed value. For the lower photon flux,
clouds of a fixed density are at smaller radius compared to the
fiducial model. At a fixed distance, the total gas masses are smaller
by about a factor of $1.5$ for the reduced flux models. For model
CHVCs at Local Group distances ($d\gtrapprox 0.5$ Mpc), the gas masses
are $\Mgas\sim 10^7-10^8\msol$. 

There is an additional constraint that we can apply to the CHVC
models. The linewidth distribution of CHVCs is sharply peaked at a
FWHM $\Delta V\approx 25\kms$ (de Heij, Braun, \& Burton 2002). We
have therefore calculated the expected line FWHM for the models (\ie
using the velocity dispersion for a self-gravitating system), both
with and without a dark matter component (in the former case the total
mass has simply been scaled up by the assumed ratio of dark to
baryonic mass); $\Delta V$ has been calculated assuming a uniform
spherical mass distribution. In Figure 4 we show the predicted line
FWHM as a function of distance, for dark matter to gas mass ratios of
ten and zero. In the former case, the models are consistent with the
observations for a distance $d\sim 330$ kpc; with no dark matter the
clouds must lie at approximately 1.5 Mpc. For $\phi_i$ reduced by a
factor of two, these distances increase somewhat, to $d\sim 430$ kpc
and 2.1 Mpc for $\Mdm/\Mgas=10$ and no dark matter, respectively.

While these models rule out the possibility that the CHVCs sample a
population of cosmological objects (\ie gas in dark matter-dominated
potential wells) at distances characteristic of the Local Group
($d\sim 1$ Mpc), they suggest that they could represent such objects
at distances of $\sim 300-400$ kpc. However, as we will see in the
next section, models of CHVCs in realistic dark matter potential wells
require that they lie much closer to the Galaxy, with an upper limit
to the distance of about 200 kpc. This distance scale rules out models
in which the CHVCs represent continuing infall onto the Local Group.

\subsection{Dark matter halo models}
Motivated by the order-of-magnitude agreement between the properties
of model clouds with $\Mdm/\Mgas\sim 10$ and the observations for
characteristic distances $d\sim 350$ kpc, we have constructed models
of CHVCs using realistic dark matter potential wells. Numerical
simulations of halo formation in cold dark matter cosmologies predict
a specific form for the density profile, as initially found by
Navarro, Frenk, \& White (1996). However, there is considerable
disagreement as to whether this profile describes the halos of real
galaxies, particularly dwarf and low-surface brightness galaxies (\eg
Flores \& Primack 1994; Moore 1994; Burkert 1995). A number of authors
have argued that these latter objects have nearly constant-density
cores rather than the cusped power-law NFW density profile, but
this point is controversial. We have therefore adopted two different
models for the halo profile. The first is the the Navarro, Frenk, \&
White (1996) halo density profile, but modified to allow for the
possible presence of a density core:
\be
\rho_{dm}(r)=\rho_{crit}{\delta_c r_s\over (r+r_o)(1+r/r_s)^2}
\ee
where $\rho_{crit}=3H^2/8\pi G$ is the closure density, the scale
length $r_s$ is related to the virial radius of the halo by the
concentration parameter $c=R_{vir}/r_s$, $r_o$ is the core radius, and
$\delta_c$ is a characteristic density contrast, determined by the
requirement that the halo mass
\be
M_h(R_{vir})=\Delta \rho_{crit}(4\pi/3)R_{vir}^3,
\ee
where $\Delta$ is the overdensity parameter. For a flat
($\Omega+\Lambda=1$) universe, $\Delta$ is given by (\eg Eke, Navarro,
\& Frenk 1998) $\Delta=178 \Omega^{0.45}\approx 111$ at $z=0$ for the
$\Omega_{dm}=0.35$ cosmology adopted here. For the relevant mass range
of halos, $c\sim 10$.

In the limit of no core, the density contrast is given by the standard
result (\eg Eke, Navarro, \& Steinmetz 2001)
\be
\delta_c ={\Delta\over 3} {c^3\over \ln(1+c)-c/(1+c)}\; .
\ee
Since we have no physical basis for choosing the core radius (this
point is discussed further below) we have kept the concentration
parameter fixed (for a given halo mass) at the no-core value, and
calculated $\delta_c$ from the requirement that the integral of
equation (1) to $R_{vir}$ equal the mass given by equation (2). The
NFW halo parameters ($c$, $R_{vir}$) for a given mass have been
calculated using the results of Mo, Mao \& White (1998) and Eke \etal\
(2001). 

The second halo mass profile is that suggested by Burkert (1995) based
on observations of dwarf galaxies,
\be
\rho_{dm}(r)={\rho_o r_o^3\over (r+r_o)(r^2+r_o^2)}
\ee
where $\rho_o$ is the core density and $r_o$ is again the core
radius. Like the NFW profile, the density of the Burkert halo falls as
$r^{-3}$ at large radius. Burkert found that the core density and
radius were tightly correlated for the observed galaxies, so that
the dark matter profiles are described by only one free parameter,
which can be taken to be $r_o$. Burkert also concluded that the halo
virial radius $R_{vir}\approx 3.4 r_o$, so that these halos are much
less centrally concentrated than the NFW halos.

We have calculated the gas density profile within these halos,
including the effects of the gas self-gravity. The dark matter density
profile is fixed, however: we do not include the response of the halo
to the gas. (This would only strengthen our conclusions on the
viability of models with dark matter.) In most cases, the gas velocity
dispersion is taken to be $\sigma_g=10.6\kms$, which produces a line FWHM
equal to the typical observed value $\Delta V=25\kms$, and is the
expected value for gas at $T\sim 10^4$ K, characteristic of the warm
neutral medium. For NFW halos, in the case of no core and negligible gas
self-gravity, the gas density profile is given by
\be
\rho_g(\tilde r)=\rho_g(0) e^{C_h\left[\ln(1+\tilde r)/\tilde r-1\right]}
\ee
where $\tilde r=r/r_s$ and 
\be
C_h={4\pi G \rho_{crit}\delta_c r_s^2\over \sigma_g^2}=7.34\times 10^{-3}
{\delta_c r_s^2\over \sigma_g^2}
\ee
for $r_s$ in kpc and $\sigma_g$ in \kms. For Burkert halos the gas
profile is not analytic even in the absence of self-gravity.

We first discuss the results of models with NFW halos, and then the
Burkert halo models.

\subsubsection{NFW halo models}
The uniform density models that are consistent with a dark matter to
baryon ratio of about 10 have total hydrogen volume densities of
approximately $3\times 10^{-3}\pcubcm$ and total gas masses of about
$M_{gas}\approx 1.4\times 10^7\msol$. We have therefore calculated the
gas distribution and the neutral fraction for a halo with a mass
$M_{dm}=1.4\times 10^8\msol$. The virial radius of this halo is 13 kpc
and the concentration parameter $c=12.44$. We have also allowed for
tidal truncation of the halo due to the Galaxy's tidal field; this
introduces a cutoff at a radius of 10.2 kpc.

In Figure 5 we show the gas density profile for the model that
satisfies the requirement that the peak $\Nhi=10^{19}\psqcm$. This
profile (which is characteristic of all the halo models) immediately
reveals the problem that plagues all of these models. Most of the gas
mass in the halo is at large radius, where the density is low and
therefore the neutral fraction is very small. Hence, in order to match
the observed neutral column density, the density at the center of the
halo must be raised substantially over that in the uniform density
model. As Figure 6 (which plots the cumulative dark matter and gas
masses as a function of radius) demonstrates, this large central gas
density results in a gas to dark matter mass ratio of
approximately unity within the halo, which is obviously untenable.
This conclusion is unaltered by reducing the ionizing flux by a factor
of two or by allowing the dark matter profile to have a core. Although
this latter modification results in gas density profiles that are more
nearly constant over a larger range in radius, and therefore produce
the same total hydrogen column density for a smaller central density,
the slower drop-off of \nh\ with $r$ due to the core means that these
models always have a higher value of $\Mgas/\Mdm$ at the outer
boundary than the no-core models. (In fact, the models with cores
invariably require larger total hydrogen columns than the models
without cores, which makes the problem even worse.)

The only way to produce reasonable baryon to dark matter ratios in
these models is to increase the mass of the halo. In order to produce
a model in which the ratio $\Mgas/\Mdm$ is less than about
0.1\footnote{Assuming that galaxy clusters provide a fair sample of
the universe, the baryon to dark matter mass ratio is estimated to be
0.13 for $H_o=70$ \kms\ Mpc$^{-1}$ (Mohr, Mathiesen, \& Evrard 1999).}
at the outer boundary {\it prior} to tidal truncation, the halo mass
must be $\Mdm\gtrapprox 5\times 10^8\msol$. The mass increase is
smaller than the factor of ten one might expect from Figure 6 because
higher mass halos are both larger than lower mass ones and denser at
the same physical radius, and therefore the observed neutral hydrogen
column can be produced with a smaller total gas mass than is required
in the lower mass halo.

In addition to the mass problem, to which we return below, these NFW
dark matter-dominated models suffer from another, severe problem for
the cosmological hypothesis: the physical size of the neutral hydrogen
distribution is so small that the model CHVCs are restricted to lie
less than $d\sim 200$ kpc from the Galaxy. This is illustrated in
Figure 7, which shows the projected \HI\ column as a function of
impact parameter for a variety of halo masses, core radii, and
ionizing fluxes. {\it In no case does the model CHVC radius at the
HIPASS threshold, $R_{18.3}$, significantly exceed 1 kpc.} As
discussed in \S 4.1, this constrains the clouds to distances of no
more than about 200 kpc from the Galaxy. In fact, except for the
lowest mass ($M_{dm}=1.4\times 10^8\msol$) model shown in Figure 7,
which, as discussed above, has an unacceptably large gas to dark
matter mass ratio, all of the models have $R_{18.3}\lessapprox 0.5$
kpc unless we include a core in the halo density distribution. As we
noted earlier, we have no physical basis for choosing the halo core
parameters, and have included a core simply to see whether, given the
uncertainties in the actual halo dark matter profiles, this could
alleviate the problems arising from the small distance scale mandated
by the small physical size of the model CHVCs. As is evident from
Figure 7, it does not. The reason is simply that if the halo is
massive enough that the observed neutral hydrogen column density can
be reproduced with a reasonable baryon to dark matter mass ratio, the
velocity dispersion of the baryonic component (which is fixed by the
observed linewidth) is only a fraction of the velocity dispersion
characterizing the dark matter potential well, with the result that
the gas is confined to the core of the halo. Increasing the size of
the core beyond $r_o\approx r_s$ does not lead to any significant
further increase in $R_{18.3}$, as is seen in the right-hand panel of
Figure 7.

\subsubsection{Burkert halo models}
In Figure 8 we show results for models with a Burkert halo of mass
$M_{dm}=10^8\msol$. For these models we have also explored a larger
region of parameters for the CHVCs, considering neutral hydrogen
columns up to ten times larger than the typical value seen in the
HIPASS survey and linewidths up to a factor of two larger than the
typical value. The left-hand panel assumes the standard velocity
dispersion of $\sigma_g=10.6$ \kms. From left to right, we plot the
neutral hydrogen column as a function of impact parameter for peak
neutral hydrogen column densities $\Nhi=10^{19}\psqcm$ (solid line),
$\Nhi=2.5\times 10^{19}\psqcm$ (long-dashed line), $\Nhi=5.0\times
10^{19}\psqcm$ (dotted line), $\Nhi=7.5\times 10^{19}\psqcm$
(short-dashed line) and $\Nhi=10^{20}\psqcm$ (solid line). This
Burkert halo model does not suffer from the unacceptably high baryon
to dark matter mass ratio problem that afflicts the NFW halos of
comparable mass, because the Burkert halo is physically much smaller
(virial radius of 1.8 kpc compared to 11.7 kpc for the same mass NFW
model) and therefore its mean density is much higher than the
corresponding NFW halo. However, this small size scale becomes a
serious problem just as it does for the NFW halos. Even for a peak
column $\Nhi=10^{20}\psqcm$, the projected size at the HIPASS sensitivity
threshold barely reaches 1 kpc, again implying that the halos are no
more than 200 kpc distant. The right-hand panel shows the same models,
only for a doubled velocity dispersion, so that the line FWHM $\Delta
V=50\kms$. Even with this assumption, the physical size of the clouds
remains small: raising the peak column density to $\Nhi=5.0\times
10^{19}$ - five times the typical HIPASS value - only increases the
cloud radius to about 1.6 kpc, which would push the clouds out to no
more than 270 kpc. (Lower mass halos - \eg $10^7\msol$ - with Burkert
profiles are so small that they can lie no further than $d\sim 100$
kpc from the Milky Way, independent of total column density.)

In Figure 9 we present similar results, but for a halo with
$M_{dm}=10^9\msol$. In the left-hand panel, with a velocity dispersion
corresponding to the typical observed value, the results scarcely
differ from those for the $10^8\msol$ halo, and the cloud size at the
HIPASS threshold again never exceeds 1 kpc. The results for the models with
twice the linewidth are also very similar to those for the lower mass
Burkert halo. The chief difference is that, since the $10^9\msol$ halo
is physically larger (by a factor of 2.7) than the $10^8\msol$ halo,
the highest column density models are larger in the higher mass model,
since the column density cutoff is not set by the physical edge of the
halo as it is in the $10^8\msol$ model. 

Hence the CHVC models with Burkert halos suffer from the same distance
problem as the NFW halo models: the sizes at the HIPASS sensitivity
threshold are simply too small to allow them to be placed at distances
much larger than $d\approx 200$ kpc from the Milky Way. Our results
are in excellent agreement with those of Kepner, Babul \& Spergel
(1997) and Kepner \etal\ (1999), who calculated the chemical, thermal,
and ionization equilibrium of gas in dark matter minihalos with
Burkert density profiles. They found that the characteristic radius at
the $\Nhi\approx 10^{18}\psqcm$ level never exceeded $r\approx 1$ kpc,
even when the central hydrogen column density reached values much
larger than any we consider here.

The combination of these two factors (the large mass that is required
in order to have plausible gas to dark matter ratios for NFW halos,
and the close distances required by the small physical size of the
clouds for either NFW or Burkert halos) is fatal for any model in
which the CHVCs represent a population of cosmological objects. In
Figure 10 we show the halo mass distribution for a cosmological
simulation of the Local Group (Moore \etal\ 2001). The $\sim 2000$
halos were created by analyzing the formation of a binary pair of
massive halos in a hierarchical universe dominated by cold dark
matter. The masses, separation and relative velocities of the binary
halos are close to those observed for the Milky Way and Andromeda. The
binary system was chosen from a large cosmological simulation such that a
nearby massive cluster similar to Virgo was present. The simulation is
described in detail in Moore \etal\ (2001).
The filled circles show all of the halos (aside
from those representing the Milky Way and Andromeda galaxies) within
the simulated Local Group volume and the open circles show the halos
that lie within 200 kpc of the Galaxy. As we showed above, the minimum
acceptable mass for CHVC halos is $M_{dm}\sim 5\times 10^8\msol$. From
this mass limit up to $M_{dm}=2\times 10^9\msol$, there are 16 halos
in the simulation\footnote{For a lower mass limit of $M_{dm}\sim
  10^8\msol$, as allowed for Burkert profile halos, the number of
  halos would be a few times larger. However, except for extreme
  assumed CHVC parameters (\ie peak column densities and linewidths
  much larger than the typical values), the CHVC models with Burkert
  halos will be physically even smaller than those with NFW
  halos. Hence these objects would have to lie closer than 200 kpc,
  and so the total number predicted would be about the same as for the
  NFW-halo models.} To explain the CHVCs as cosmological, dark
matter-dominated objects, we need approximately 250. Hence unless the
hierarchical, cold dark matter-dominated cosmology grossly {\it
under-predicts} the number of halos in this mass range, our results do
not allow CHVCs to represent such a population.

\subsection{CHVC HI distribution}
There have now been several detailed studies of the HI structure of
CHVCs. Wakker \& Schwarz (1991) investigated two CHVCs and were the
first to notice the core/halo structure of the clouds. Braun \& Burton
(2000) imaged six CHVCs at 1$^{\prime}$ resolution and found compact
cores with linewidths as low as 2 - 10 \kms, as well as the large
diffuse halo modeled here with linewidths of $\sim 25$ \kms.  The warm
($\sim 10^4$ K) halos were imaged at 3$^{\prime}$ resolution with
Arecibo by Burton, Braun \& Chengalur (2001; hereafter BBC).\footnote{
Of the ten CHVCs imaged by BBC at Arecibo, 8 are covered by the
northern extension of the HIPASS survey. We have examined the HIPASS
data for these clouds to see whether the HIPASS results have been
seriously compromised by resolution effects. With the exception of the
faintest cloud in the sample (for which the Arecibo peak column is
about 4.5 times the HIPASS value: however, the column density peaks
are due to probably unresolved, non-centered structure, and the entire
western half of the cloud has a typical column extremely close to the
HIPASS value), the Arecibo and HIPASS peak column
densities differ by at most a factor of two. Hence there is no reason
to think that the HIPASS observations systematically underestimate the
cloud column densities significantly enough to affect the results of
this paper.}
They argued that the column density distribution at the edges of the
clouds (for \Nhi\ $ < 10^{18.5}$ \psqcm) drops off as an exponential
with radius, indicating a 
spherical exponential distribution of neutral hydrogen volume density
as a function of radius. (Figures 11 and 12 of BBC demonstrate the
quality of the exponential fits.) 
Kinematically the cores and halos of the CHVCs do not
appear to be related. BBC derived distances of typically a few hundred
kpc for the clouds in their sample, based on the following
assumptions:
\begin{itemize}
\item The neutral gas density distribution is reasonably
  well-described by an exponential with radius, so that the central
  density can be inferred from the peak column density and the fit
  scalelength; 
\item The gas neutral fraction is close to unity in the center of the
  CHVC, and the temperature $T\sim 10^4$ K;
\item The total gas pressure at the center is $P/k\sim 100$ cm$^{-3}$
  K, which Braun \& Burton (2000) argued is the characteristic
  pressure at which the cold cores can co-exist with the warm neutral
  medium. 
\end{itemize}
With these assumptions, the angular scale length of the exponential
  can be converted to a physical scale length, thereby yielding a
  cloud distance. 

In this section we show that the characteristic exponential
\Nhi\ profile arises as a simple consequence of the physics of
photoionization, and in the absence of fine-tuning of conditions, is
very unlikely to provide reliable distance estimates.

In Figure 11 we plot the projected neutral hydrogen column density as a
function of impact parameter (normalized to the cloud radius) for five
different uniform density models, with total densities \nh\ between
$10^{-4}$ and $10^{-2}\psqcm$. For all of these models the \Nhi\
profile resembles an exponential over most of the cloud radius, with
the exception of a core which becomes more pronounced with increasing
density, and of a cutoff as the cloud outer boundary is
approached. The exponential appearance of the \HI\ column density
distribution indicates that the neutral hydrogen volume density
distribution is also approximately an exponential, at least over a
substantial fraction of the cloud volume. This is simply a consequence
of the photoionization physics, and reflects the attenuation of the
ionizing radiation (and therefore the increase in the neutral
fraction) with depth into the clouds; the total gas volume densities
in all these models are uniform. 

BBC fit \Nhi\ profiles of the form
\be 
\Nhi(p)=2 n_o h\left({p\over h}\right)K_1(p/h)
\ee
to their observations, where $p$ is the impact parameter, $n_o$ is the
central density, $h$ is the scale-length, and $K_1$ is a modified
Bessel function. This is the form expected for an exponential volume
density distribution of infinite extent (van der Kruit \& Searle
1981). This is modified for a cloud of finite extent, however, and so
instead we use the finite-cloud profile
\be
\Nhi(p)=2 n_o h\left({p\over h}\right)\left[e^{-R/h}
\left(R^2/p^2-1\right)^{1/2}+ k_1\left(p/h,R/p\right)\right]
\ee
where $R$ is the cloud radius and $k_1$ is an incomplete Bessel
function,
\be
k_1(\rho,x)=\rho\int_1^x e^{-\rho t}\left(t^2-1\right)^{1/2}\,dt
\ee
which goes to the usual $K_1(\rho)$ as $x\rightarrow\infty$.

In Figure 12 we show two examples of fits of equation (8) to the
projected neutral hydrogen column density distributions for the CHVC
models described in \S 4.1. These are not formal fits to the
distributions, and are intended solely to show that these uniform
density models are very well described over a large range in radius
(in fact, everywhere except for the cores) by an exponential \nhi\
distribution. Although BBC argued based on observations of dwarf
galaxies that there is a characteristic value of $h$, that conclusion
does not hold for these models, as $h$ depends nontrivially on the
volume density. For the pair of models shown in Figure 12, which
differ in \nh\ by a factor of two, the fitted values of $h$ differ by
nearly a factor of four.

The BBC distance method generally yields erroneous results when
applied to these models, for two reasons: (1) many of the CHVC models
have substantial ionized fractions even in their cores, so that the
central neutral density - even if determined with reasonable accuracy
from the exponential fit - can be substantially below the actual total
gas density, and (2) the central pressure is generally not equal to
$P/k\sim 100$ cm$^{-3}$ K. \footnote{For the $\nh=0.01\pcubcm$ model,
for which $P/K$ does equal 100 at cloud center, the BBC distance
method gives a result within about $25\%$ of the value assigned to the
cloud ($D=67$ kpc, Table 1) in order to make its angular size match
the typical HIPASS size, provided that the true central column density
is used, rather than the fitted column density, since in these
relatively high-density models the presence of a core means that the
exponential fit overestimates the central column density - \cf Figures
11 and 12.} It is not at all obvious that the value of $P/k\sim 100$
quoted by Braun \& Burton (2000) for the co-existence of cold gas with
the warm neutral medium is generally applicable. This conclusion was
based on a phase diagram calculated by Wolfire \etal\ (priv. comm.) for
``Local Group conditions'', which are not otherwise specified. In the
calculations of Wolfire \etal\ (1995a,b), which presumably use similar
physics, gas heating is dominated by grain photoelectric heating, and
the only source of ionizing photons is the extragalactic soft X-ray
background, which produces an ionizing flux about an order of
magnitude smaller than assumed here. Furthermore, the resulting phase
diagram is rather sensitive to the adopted gas parameters (see Figures
1 and 6 in Wolfire \etal\ 1995b, in which the range of pressures over
which a two-phase medium is allowed can extend over three orders of
magnitude). In fact, in the phase diagram presented by Braun \& Burton
(their Figure 13), the minimum pressure $P_{min}$ for which the cold
phase exists is approximately five times larger ($P_{min}/k\approx 300$
cm$^{-3}$ K) for a CHVC warm gas column density $\Nhi=10^{19}\psqcm$,
as appropriate for the HIPASS CHVCs discussed here, than it is for
$\Nhi=10^{20}\psqcm$ (for which $P_{min}/k$ is actually about 60), as
adopted by Braun \& Burton for their sample. This would reduce the
inferred distances by a factor of three. It is also not yet clearly
established that the presence of cold gas is a generic property of
CHVCs: there are only six clouds in the Braun \& Burton (2000) sample,
and while all of these exhibit cold cores, a much larger sample is
clearly needed. The Magellanic Stream shows no evidence for cold gas
(Mebold \etal\ 1991).

Interestingly, the models in which the gas is confined by a dark
matter potential are not well described by profiles of the form (7) or
(8). We show one such NFW model in Figure 13. This is the $\Mdm=5\times
10^8\msol$, no-core model shown as the solid line in the center panel
of Figure 7. Inspection of that figure shows that the shape of the
neutral hydrogen column density distribution in these dark
matter-dominated models is generic. As is evident from Figure 13,
these models cannot be described by a single component of the form (7)
or (8). This is because the neutral hydrogen column density always
exhibits a core with an extended wing, in consequence of the gas
density profiles in these halos (\cf Figure 5). Hence the profile fit
to the core of the \Nhi\ distribution (we have used equation (7) for
these models, since the models are actually much larger than the
plotted region, and hence the cutoff is unimportant here), shown by
the dashed line in Figure 13, grossly under-predicts \Nhi\ outside
$r\sim 0.5$ kpc, while the fit to the column density distribution at
larger radius (dotted line) falls an order of magnitude short of \Nhi\
at $r=0$. The results for Burkert profile halos are similar; in Figure
14 we show the neutral column density distribution for the largest
column density model ($\Nhi(0)=1.0\times 10^{20}$ \psqcm) shown
in the lefthand panel of Figure 8 ($\Mdm=10^8\msol$, $\sigma_g=10.6$
km s$^{-1}$). As for the NFW halo model, there is no reasonable fit
with a single component of the form (7) or (8); in fact, the core-halo
structure is even more pronounced than in the NFW halo. As before,
fits to the core of the \Nhi\ distribution (here we had to use
equation [8]) severely underestimate \Nhi\ beyond $r\sim 0.8$ kpc,
while fits to \Nhi\ at large radius underestimate the column density
at small radii by an order of magnitude. 

\section{Summary and Discussion}
Models in which the compact high-velocity clouds (CHVCs) lie at
distances of hundreds of kpc to $\sim$1 Mpc from the Galaxy have such
low volume densities of hydrogen that exposure to the extragalactic
ionizing background radiation alone will largely ionize them. For
models of uniform density clouds (\S 4.1), we find that the total
hydrogen column needed to produce the typical neutral hydrogen column
of \Nhi $= 10^{19} \psqcm$ ranges between 3 and 200 times \Nhi\
(Figure 1) as the total hydrogen density varies between $10^{-2}$ and
$10^{-4} \pcubcm$. The total gas masses of these clouds are therefore
much larger than the apparent neutral gas masses. To match the typical
angular size seen for CHVCs in the HIPASS survey, low-density clouds
must be at large distances and have correspondingly large
masses. Models in which the cloud hydrogen densities are less than
$\nh\approx 2-3\times 10^{-3}\pcubcm$ are ruled out, as such clouds
would have to lie outside the Local Group (\cf Figures 2 and 3).

At Local Group distances, $d\sim 0.7-1$ Mpc, the gas mass alone of the
individual CHVCs would be $\Mgas\sim 10^8\msol$, and the mass of the
observed population would be $M\sim 4\times 10^{10}\msol$. Such CHVCs
could have dark matter/gas mass ratios of only $\sim 2-3$ without
producing line widths larger than the observed value. Models in which
the CHVCs are at smaller distances, $d\sim 350$ kpc, with gas masses
$\Mgas\sim 10^7\msol$, appear to be more reasonable, and could be
dynamically compatible with plausible ($\sim 10$ to 1) dark matter to
gas mass ratios.

A closer examination of this scenario, however, using realistic dark
matter halos, shows that it runs into severe difficulties (\S
4.2). This is because: (1) the sizes of the detectable \HI\ clouds are
too small ($R\lessapprox 1$ kpc), which requires that the CHVCs are at
distances $d\lessapprox 200$ kpc (this is true whether we use NFW
profiles or Burkert profiles for the halo dark matter density
distribution); and (2) the halo masses required for 
reasonable ratios of $\Mdm/\Mgas$ for NFW halos are too large ($\Mdm\sim
10^9\msol$). Hence the predicted number of such objects in a
hierarchical CDM structure formation scenario is more than an order of
magnitude smaller than the observed number of CHVCs, and CHVCs cannot
represent a population of infalling, cosmological objects.

We have also examined the suggestion that the approximately
exponential \Nhi\ profiles seen in high resolution studies of a sample
of CHVCs could provide an estimate of the cloud distances (\S 4.3).
Such profiles are a natural consequence of an external source of
ionizing photons for clouds with uniform total density. The derived
scale lengths are not constant and depend on the cloud density. We
argue that this technique cannot be used to derive distances to the
CHVCs in a model-independent way, as the scale lengths are not fixed
and the alternative assumption of a constant pressure for the CHVC
population is debatable. The CHVC \Nhi\ profiles can be explained by
photoionization even if the clouds are near to the Galaxy
($d\lessapprox 100$ kpc).

Our models have assumed that the gas is smoothly distributed, and
therefore do not include substantial clumping. This is arguably a
plausible assumption for the CHVC envelopes that we have modeled in
this paper, as the gas in these envelopes presumably is in the warm
($T\sim 10^4$ K, as is indicated by the observed linewidths) neutral
phase, and is supported by the generally smooth distributions and
reflection symmetry seen in high-resolution studies of the envelopes
(BBC). The only way in which this assumption can be substantially in
error, and the gas highly clumped, is if the clumps are in the cold
neutral phase, with temperatures of order 100 K. In this case the
observed linewidths must reflect bulk motion rather than thermal
velocities. Since the (thermal) linewidths of individual clumps would
be $\Delta V\sim 2 \kms$, a large number of clumps would need to be
present along any given sightline in order to explain the smoothness
of the velocity profiles. Such clumps would also have to be
substantially smaller than the spatial resolution of the high
resolution imaging studies, which frequently do find (and resolve)
small numbers of cores with narrow linewidths (\eg BBC). If pressures
$P/k\sim 100\pcubcm$ K are necessary to drive the gas into the cold
neutral phase (BBC), then the hydrogen densities must
be at least $\nh\sim 1\pcubcm$. It is rather difficult to see how to
make such a density structure compatible with the observations. If the
clumps are self-gravitating, then they must have column densities that
are much larger than the beam-averaged column densities measured
observationally, which requires that the areal filling factor is small
even though the filling factor in velocity space must be large to
reproduce the observed line profiles. If they are not
self-gravitating, we are left with the question of what confines them
(or generates them, if they are transient objects). Hence we consider
it unlikely that the CHVC envelopes that we have considered in this
paper actually consist of dense, cold clumps rather than relatively
smooth warm neutral gas.

What possibilities does this leave us with for the nature of CHVCs?
Clouds without dark matter located within the confines of the Local
Group are not massive enough to be self-gravitating, and must be
either transient or confined by some external medium. Either of these
scenarios would seem to argue for the CHVCs being associated with the
Galaxy rather than the Local Group, as both the relatively short
dynamical timescale ($t\sim 10^8$ years) and the presence of a
confining medium are much more easily understood if the CHVCs are
connected to the flux of mass and energy through the Galactic halo.
This suggestion is supported by the overall similarities of the CHVCs
to the remainder of the HVC population (Putman \etal\ 2002). The total
and neutral hydrogen gas masses of the entire observed CHVC population
($\sim 250$ objects) would be $\Mgas\sim 7\times 10^8\msol$ and
$\Mhi\sim 7\times 10^7\msol$, respectively, at $d\approx 200$ kpc, and
scale nearly as $d^2$ for smaller distances.

\acknowledgements
We are grateful to several colleagues, especially Nick Gnedin and
Jessica Rosenberg, for helpful discussions. Nick kindly provided the
software (written by Oleg Gnedin) for calculating the properties of
NFW halos. PRM is supported by the National Science Foundation
under grant AST 99-00871 and by NASA through HST grant AR-08747.02-A. 
MEP is supported by NASA through Hubble Fellowship grant
HST-HF-01132.01-A awarded by the Space Telescope Science Institute,
which is operated by the Association of Universities for Research in
Astronomy, Inc., for NASA, under contract NAS 5-26555.


\newpage

\clearpage
\setcounter{figure}{0}
\clearpage
\begin{figure}
\centerline{\psfig{figure=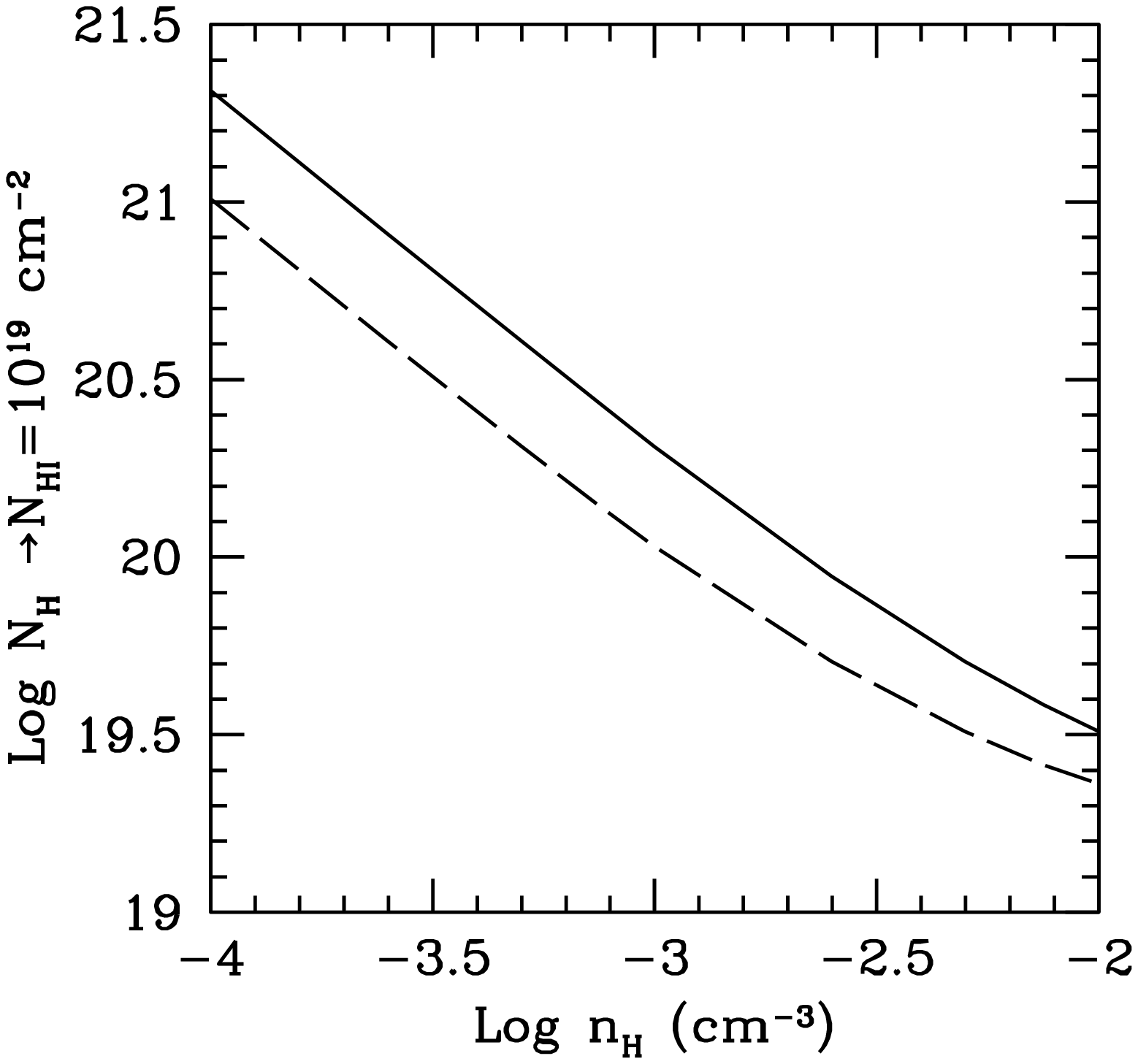,width=\textwidth,angle=0}}
\caption{The total hydrogen column density \Nh\ required to obtain a
neutral hydrogen column density \Nhi $= 10^{19}$ \psqcm, as a function
of the total hydrogen density. The solid curve is for an ionizing flux
$\phi \sim 10^4$ \phoflux; the dashed curve shows the effect of
reducing $\phi$ by a factor of two. In the latter case the values of
\Nh\ are smaller by a factor of two at the low-density end compared to
the fiducial model.}
\end{figure}

\clearpage
\begin{figure}
\centerline{\psfig{figure=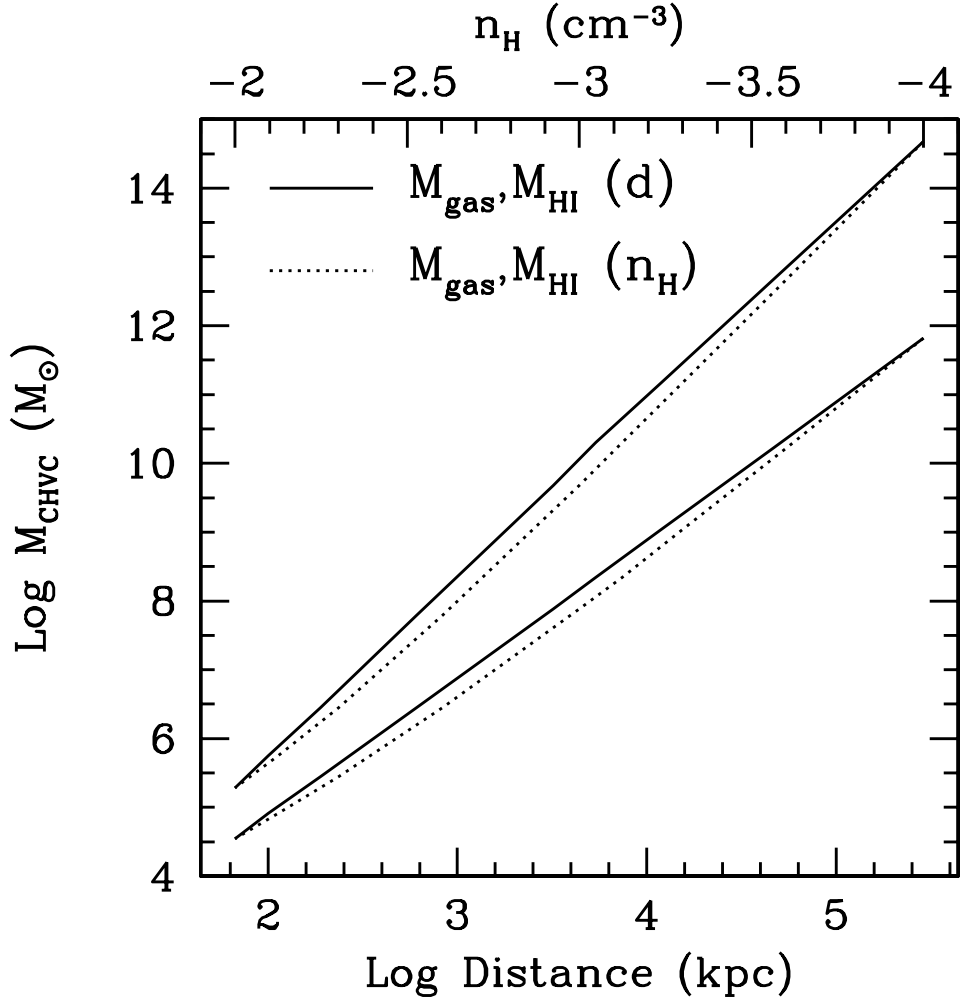,width=\textwidth,angle=0}}
\caption{The total (neutral plus ionized) and apparent (derived from
the neutral column density and apparent size) gas masses, plotted
vs. distance and total H density. For each model (\ie assumed total
hydrogen density), the distance $d$ is determined from the requirement
that the apparent angular size (determined using the radius for which
the projected neutral hydrogen column drops to the HIPASS sensitivity
limit, \Nhi $= 2 \times 10^{18}$ \psqcm) matches the typical size seen
in the HIPASS sample of CHVCs, for which the angular radius $\Delta
\theta \sim 0.34$ degrees. The masses as a function of density (top
axis) are plotted as dashed lines, and as a function of distance
(bottom axis) as solid lines; the true gas mass $M_{gas}$ is always
greater than the apparent mass \Mhi.}
\end{figure}

\clearpage
\begin{figure}
\centerline{\psfig{figure=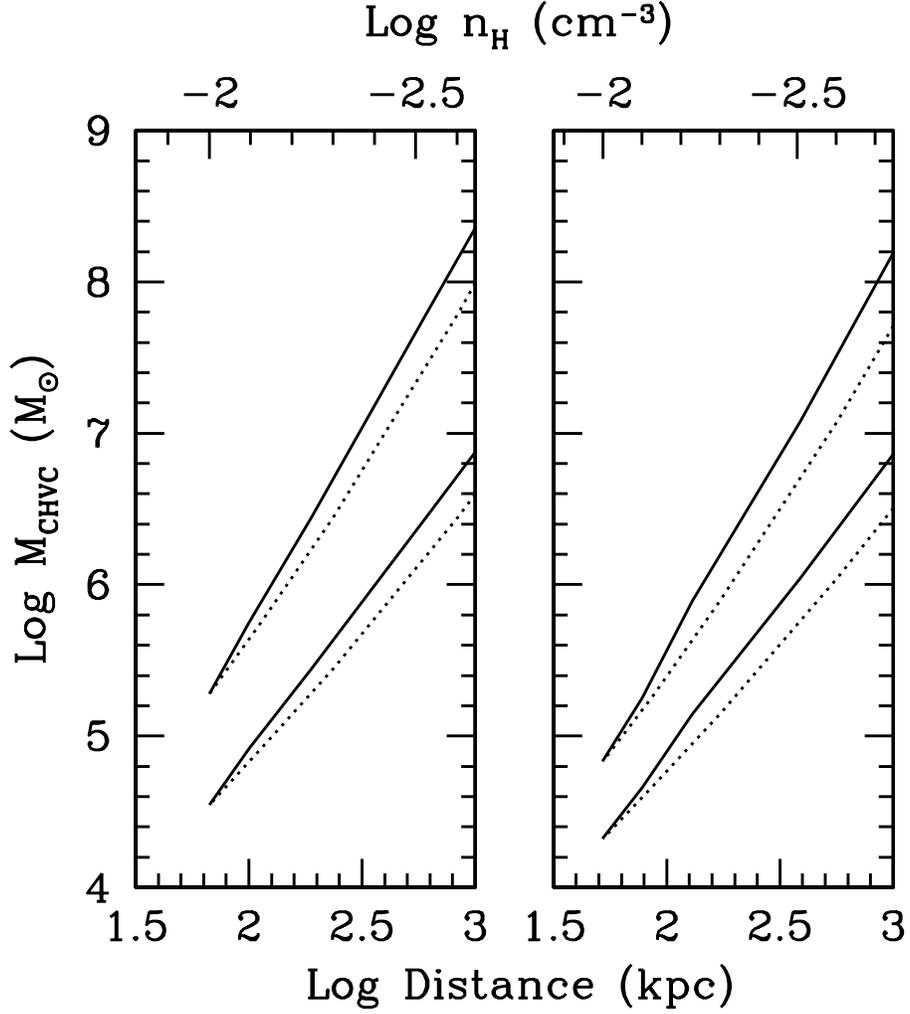,width=\textwidth,angle=0}}
\caption{The total and apparent gas masses, plotted vs. distance and
total H density, as in Figure 2. The left panel shows the results for
our fiducial ionizing flux $\phi_i\sim 10^4$; the right panel is for
$\phi_i$ reduced by a factor of two.} 
\end{figure}

\clearpage
\begin{figure}
\centerline{\psfig{figure=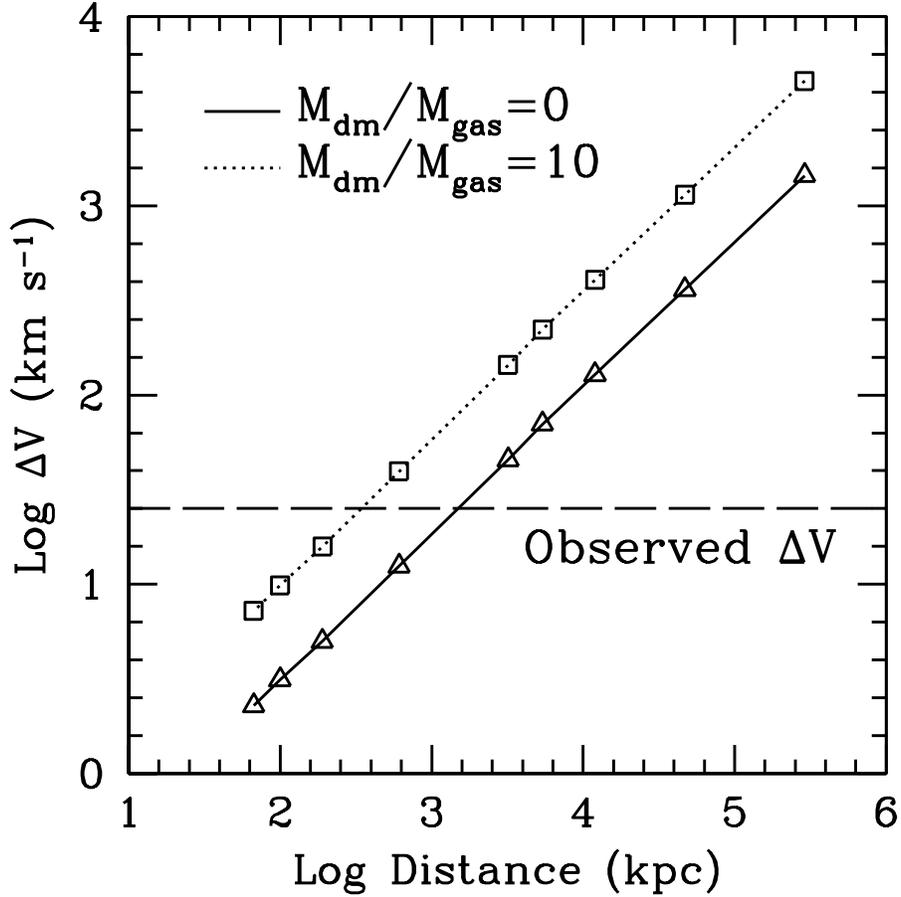,width=\textwidth,angle=0}}
\caption{Predicted line FWHM vs. distance, for dark matter to gas mass
ratios of 10 and 0, for the model CHVCs. As in Figure 2, the distance
is determined by requiring that the apparent angular size equals the
typical CHVC angular size seen in the HIPASS survey. The observed line
width, $\Delta$V $\sim 25$ \kms, is shown as the long-dashed
line. The models are consistent at $d\sim$ 330 kpc with $\Mdm/\Mgas=10$
(dotted line) and at 1.5 Mpc with no dark matter (solid line). For the
reduced $\phi_i$ models (not shown), the intercepts are at $d\sim 430$
kpc and $d\sim 2.1$ Mpc for 10:1 and no dark matter, respectively.}
\end{figure}

\clearpage
\begin{figure}
\centerline{\psfig{figure=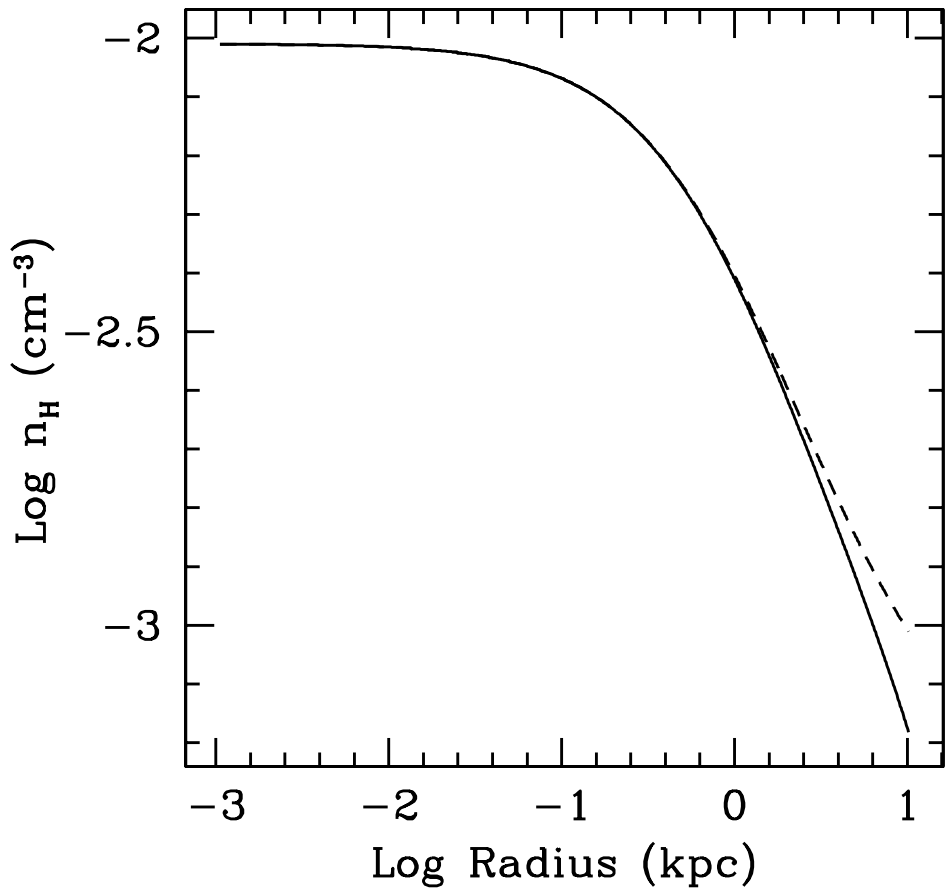,width=\textwidth,angle=0}}
\caption{The gas density distribution in a NFW dark matter halo of mass
  $M_{dm}=1.4\times 10^8\msol$ (solid line); the dashed line shows the
  density profile in the limit of negligible gas self-gravity.}
\end{figure}

\clearpage
\begin{figure}
\centerline{\psfig{figure=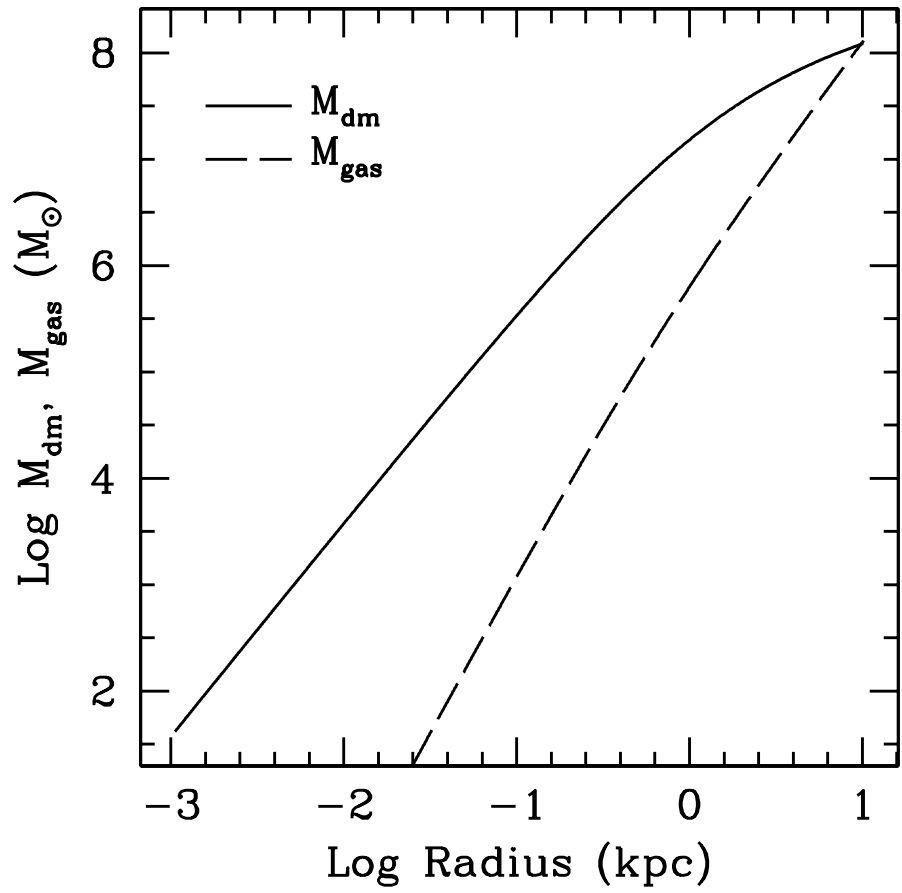,width=\textwidth,angle=0}}
\caption{The cumulative dark matter and gas masses for the halo shown
in Figure 5. The ratio $\Mgas/\Mdm\approx 1$ for this model.}
\end{figure}

\clearpage
\begin{figure}
\centerline{\psfig{figure=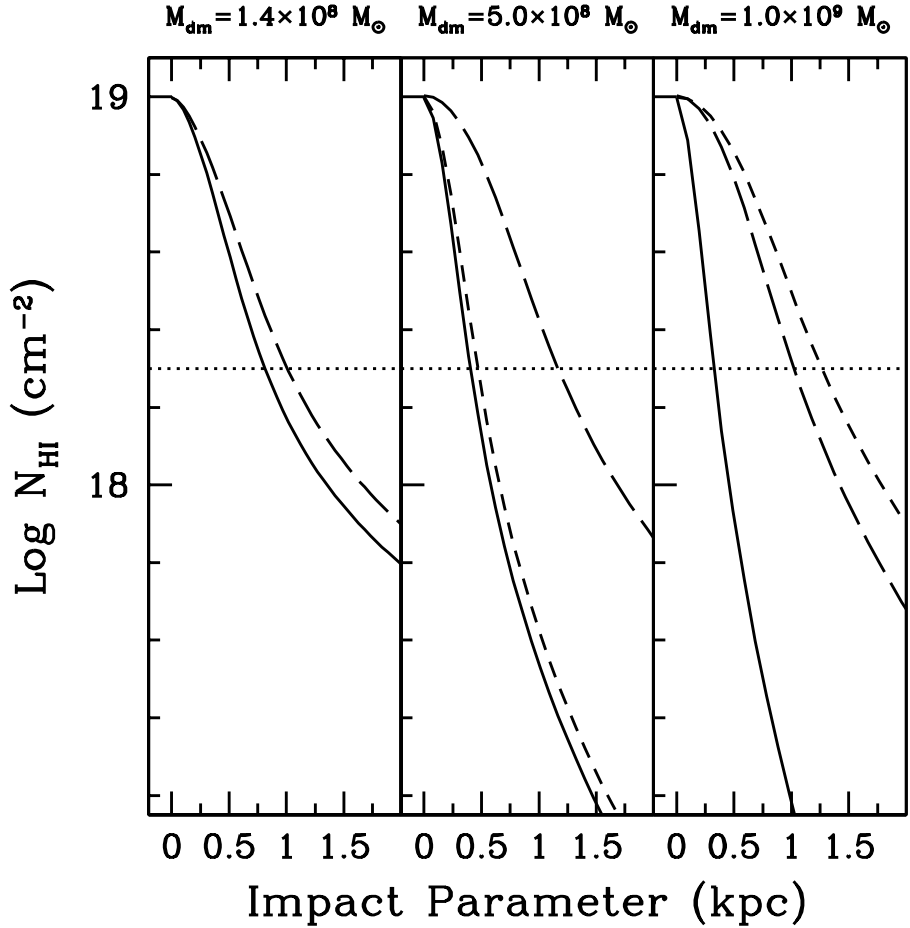,width=\textwidth,angle=0}}
\caption{The projected neutral hydrogen column as a function of impact
parameter, for several different NFW halo models. From left to right, the
panels are for halo dark matter masses $M_{dm}=1.4\times 10^8\msol$,
$5.0\times 10^8\msol$, and $1.0\times 10^9\msol$, respectively. The
dotted line running across the panels shows the HIPASS column density
threshold of $\Nhi=2\times 10^{18}\psqcm$. Several models are shown
in each panel. In the lefthand panel ($M_{dm}=1.4\times 10^8\msol$),
the solid curve is for our fiducial ionizing flux $\phi_o$, and the
long-dashed curve is for $\phi_i=\phi_o/2$. In the center panel
($M_{dm}=5.0\times 10^8\msol$), the solid line is for $\phi_i=\phi_o$,
the short-dashed line is for $\phi_i=\phi_o/2$, and the long-dashed
line is for $\phi_i=\phi_o$ and a halo core radius $r_o=r_s$. In the
righthand panel ($M_{dm}=1.0\times 10^9\msol$), all models have
$\phi_i=\phi_o$, and the solid, long-dashed, and short-dashed lines
are for $r_o=0$, $r_s$, and $2r_s$, respectively.}
\end{figure}

\clearpage
\begin{figure}
\centerline{\psfig{figure=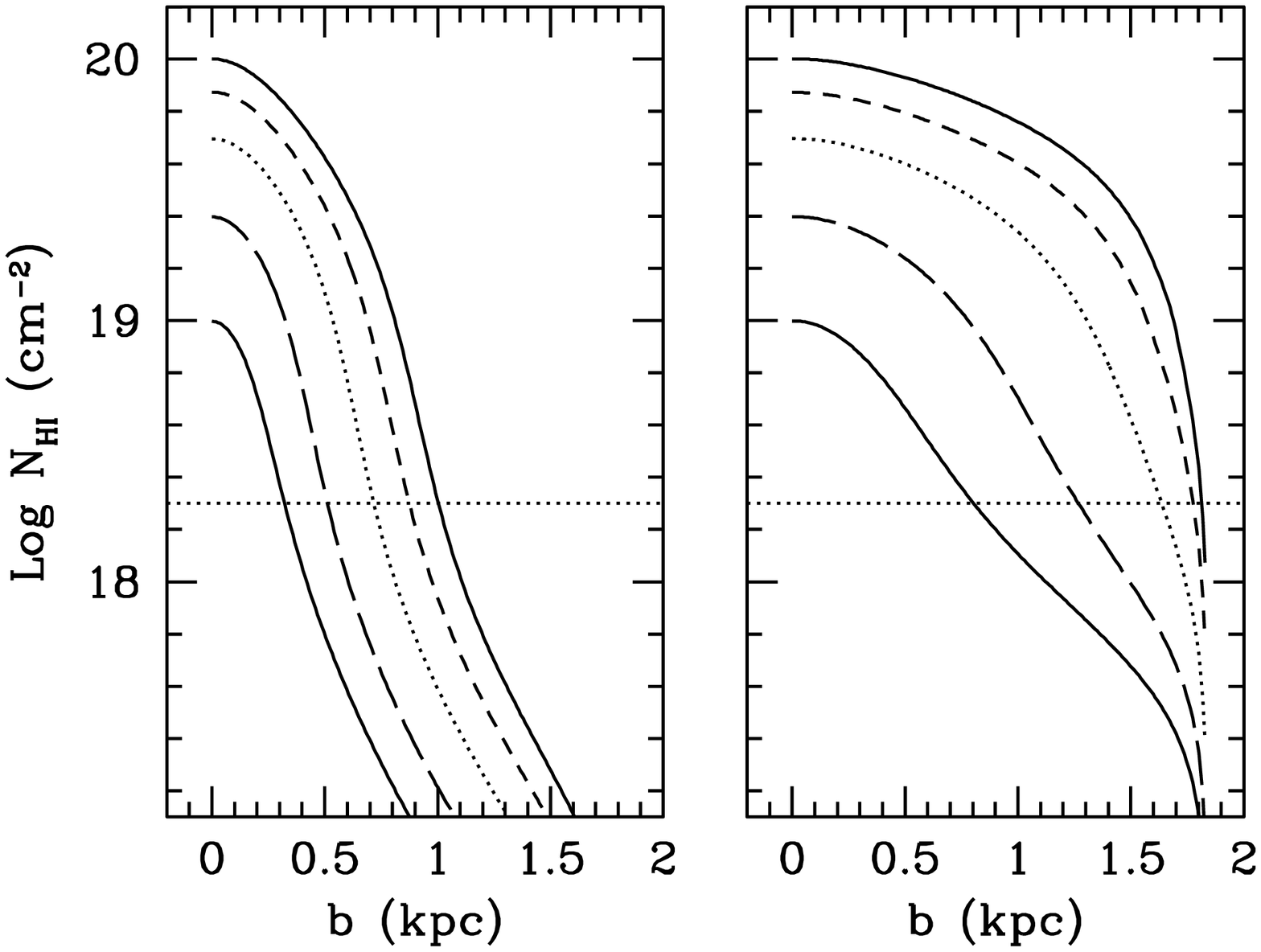,width=\textwidth,angle=0}}
\caption{The projected neutral hydrogen column as a function of impact
parameter, for Burkert halo models. The halo mass is $M_{dm}=1.0\times
10^8\msol$. From left to right, the curves are for central neutral
hydrogen column densities $\Nhi(0)=1.0\times 10^{19}$, $2.5\times
10^{19}$, $5.0\times 10^{19}$, $7.5\times 10^{19}$, and $1.0\times
10^{20}\psqcm$, respectively. In the lefthand panel a velocity
dispersion of $\sigma_g=10.6\kms$ has been assumed, while in the
righthand panel $\sigma_g=21.2\kms$. The dotted line running across
the panels shows the HIPASS column density threshold of $\Nhi=2\times
10^{18}\psqcm$.}
\end{figure}

\clearpage
\begin{figure}
\centerline{\psfig{figure=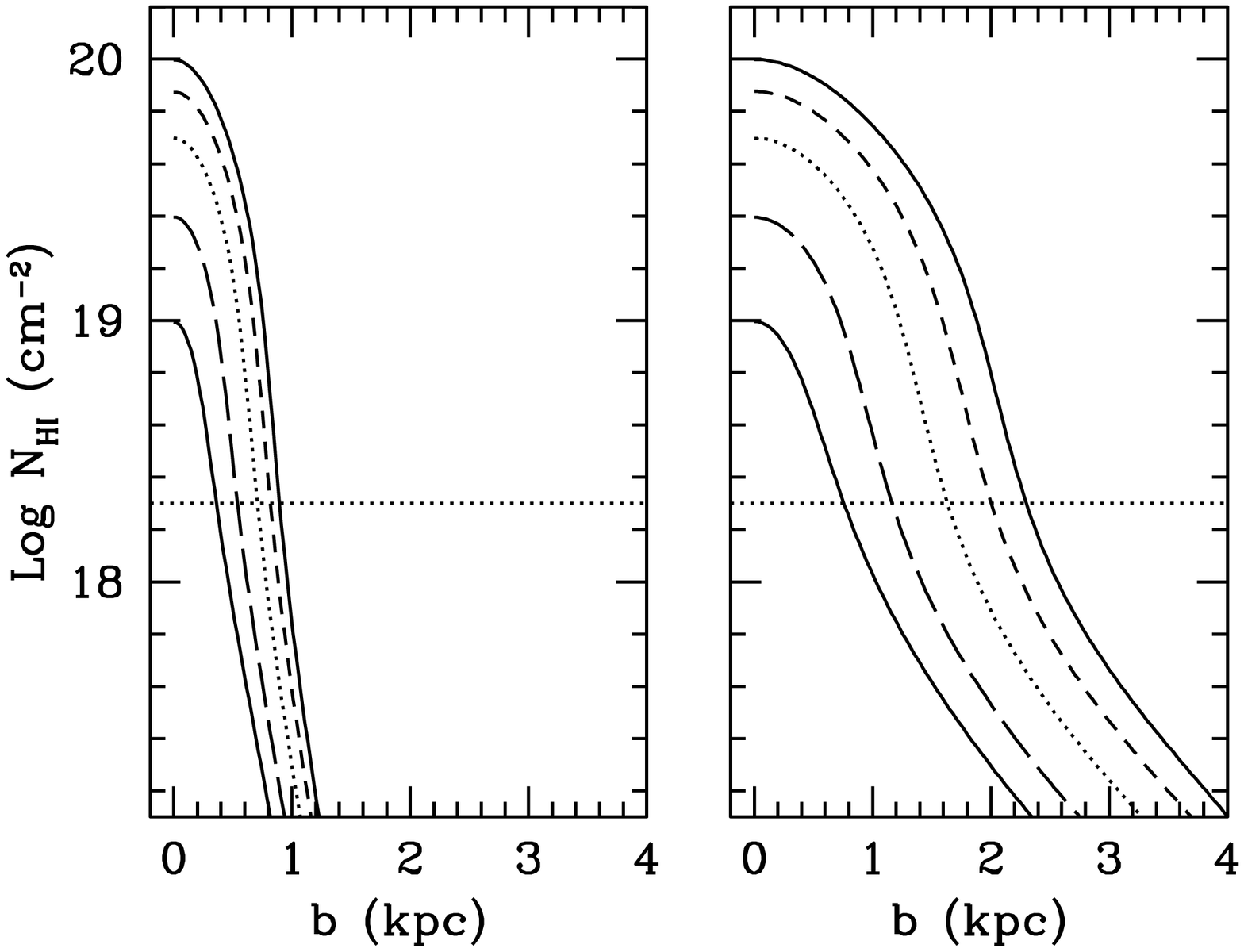,width=\textwidth,angle=0}}
\caption{The projected neutral hydrogen column as a function of impact
parameter, for Burkert halo models. The halo mass is $M_{dm}=1.0\times
10^9\msol$. From left to right, the curves are for central neutral
hydrogen column densities $\Nhi(0)=1.0\times 10^{19}$, $2.5\times
10^{19}$, $5.0\times 10^{19}$, $7.5\times 10^{19}$, and $1.0\times
10^{20}\psqcm$, respectively. In the lefthand panel a velocity
dispersion of $\sigma_g=10.6\kms$ has been assumed, while in the
righthand panel $\sigma_g=21.2\kms$. The dotted line running across
the panels shows the HIPASS column density threshold of $\Nhi=2\times
10^{18}\psqcm$.}
\end{figure}

\clearpage
\begin{figure}
\centerline{\psfig{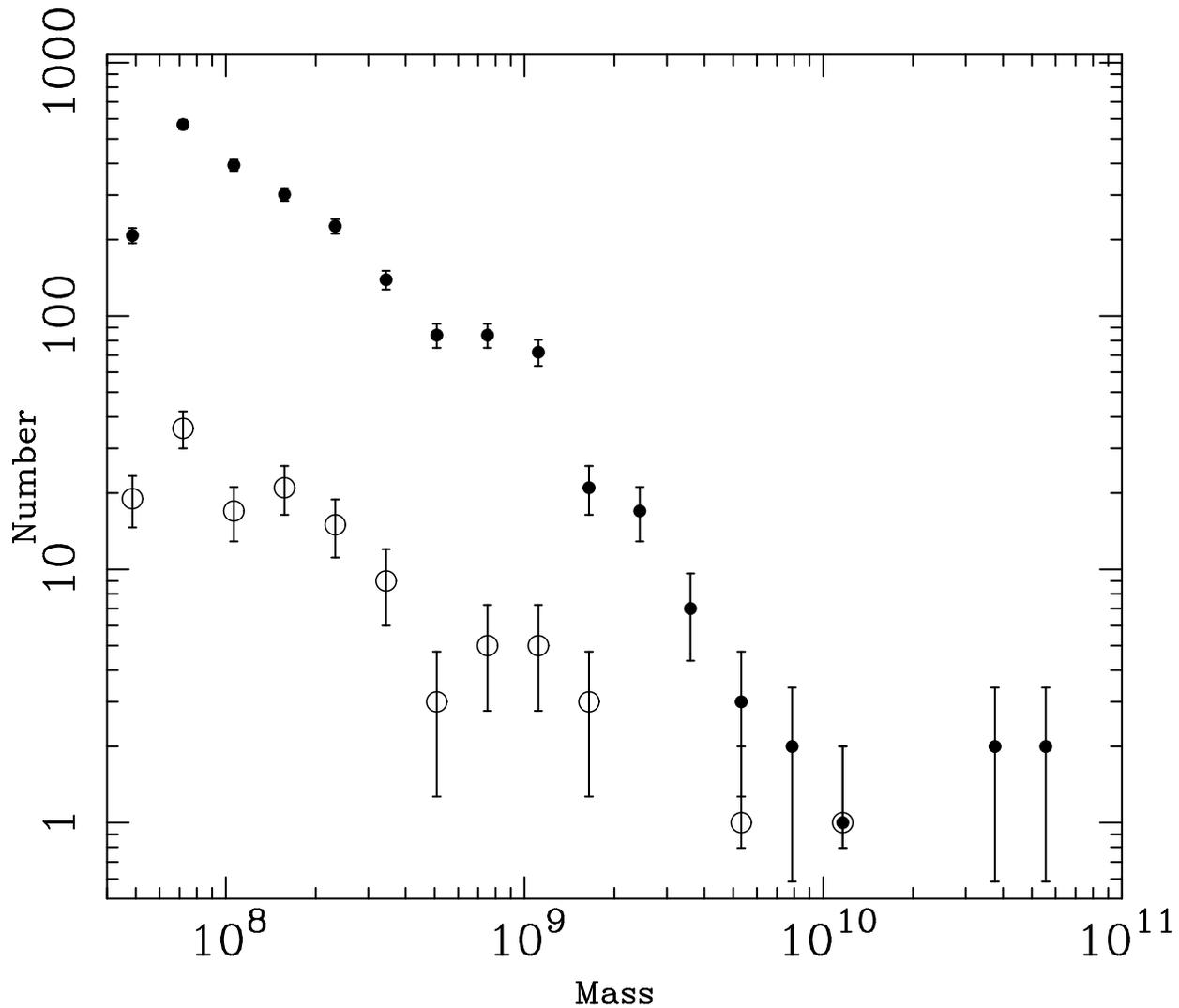}}
\caption{The halo mass distribution for a realization of the Local
 Group in a cold dark matter-dominated cosmological simulation. The
 filled circles show all of the halos (other than those representing
 the Milky Way and Andromeda) within the Local Group, and the open
 circles show all those that lie within 200 kpc of the Milky Way.}
\end{figure}

\clearpage
\begin{figure}
\centerline{\psfig{figure=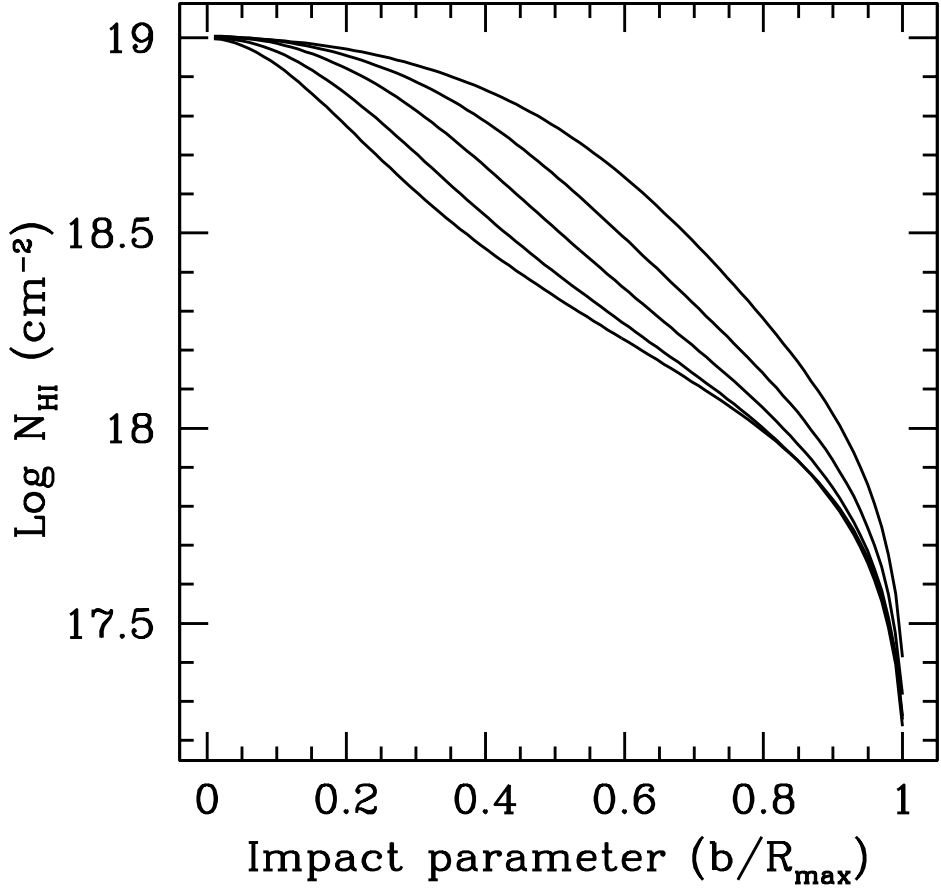,width=\textwidth,angle=0}}
\caption{The neutral hydrogen column density as a function of
normalized impact parameter (\ie the radial offset divided by the
cloud radius) for different values of total hydrogen density. From top
to bottom, the curves are for $\nh =1.0\times 10^{-2}$, $5.0\times
10^{-3}$, $2.5\times 10^{-3}$, $1.0\times 10^{-3}$, and $1.0\times
10^{-4}$ \pcubcm. The profiles generally resemble exponentials except
for the presence of a core, which becomes more prominent as the
density is increased. The profiles also show a cutoff close to the
outer boundary, but the latter occurs at neutral hydrogen columns
below the detection threshold of the HIPASS survey.}
\end{figure}

\clearpage
\begin{figure}
\centerline{\psfig{figure=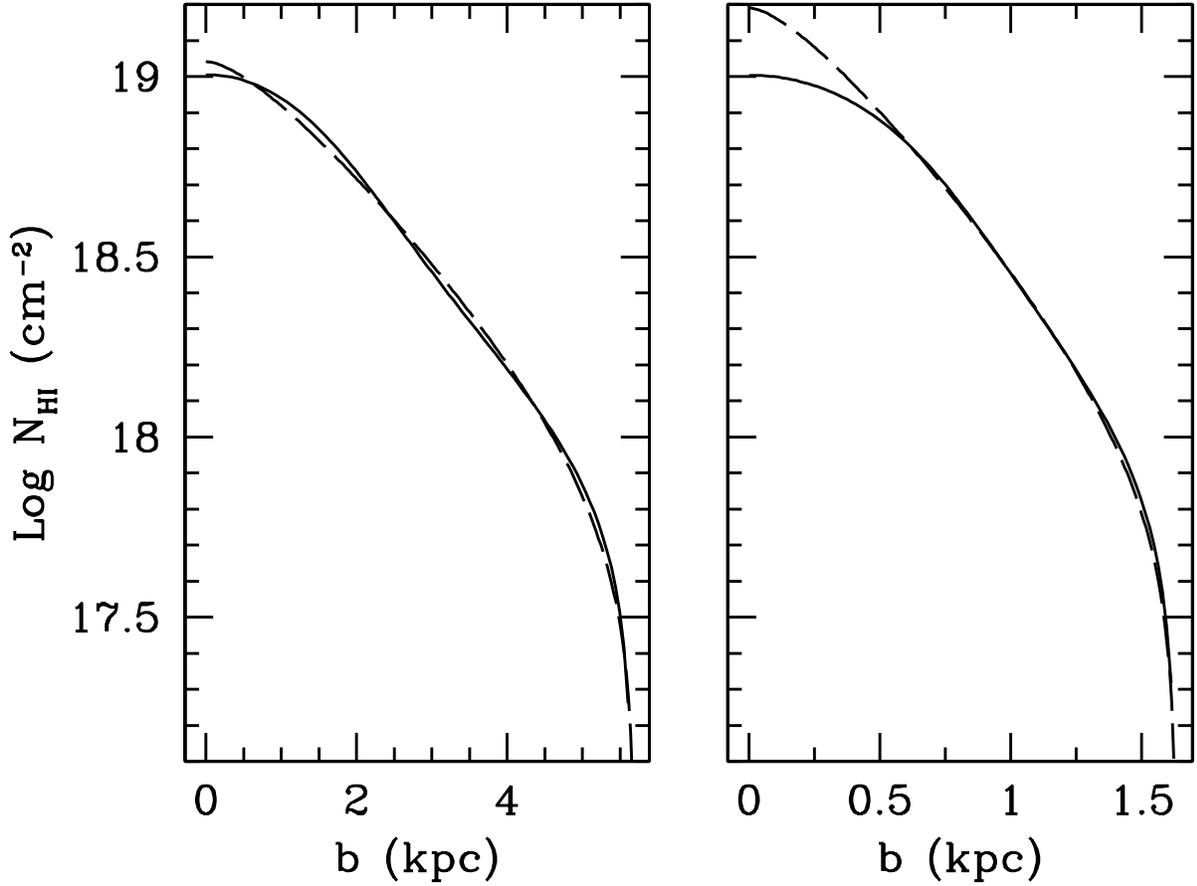,width=\textwidth,angle=0}}
\caption{The neutral hydrogen column density (solid lines) as a
function of impact parameter $b$ for uniform density clouds with
$\nh=2.5\times 10^{-3}$ (left panel), and $5.0\times 10^{-3}$ \pcubcm\
(right panel). Also plotted are fits for projected exponential volume
density distributions of neutral hydrogen, as discussed in the text
(dashed lines). In both cases the cloud radius is identical to the
maximum impact parameter. For the lefthand panel, the scale length
$h=1.5$ kpc and the central column density $\Nhi(0)=1.1\times
10^{19}\psqcm$, while for the righthand panel $h=0.42$ kpc,
$\Nhi(0)=1.55\times 10^{19}\psqcm$. }
\end{figure}

\clearpage
\begin{figure}
\centerline{\psfig{figure=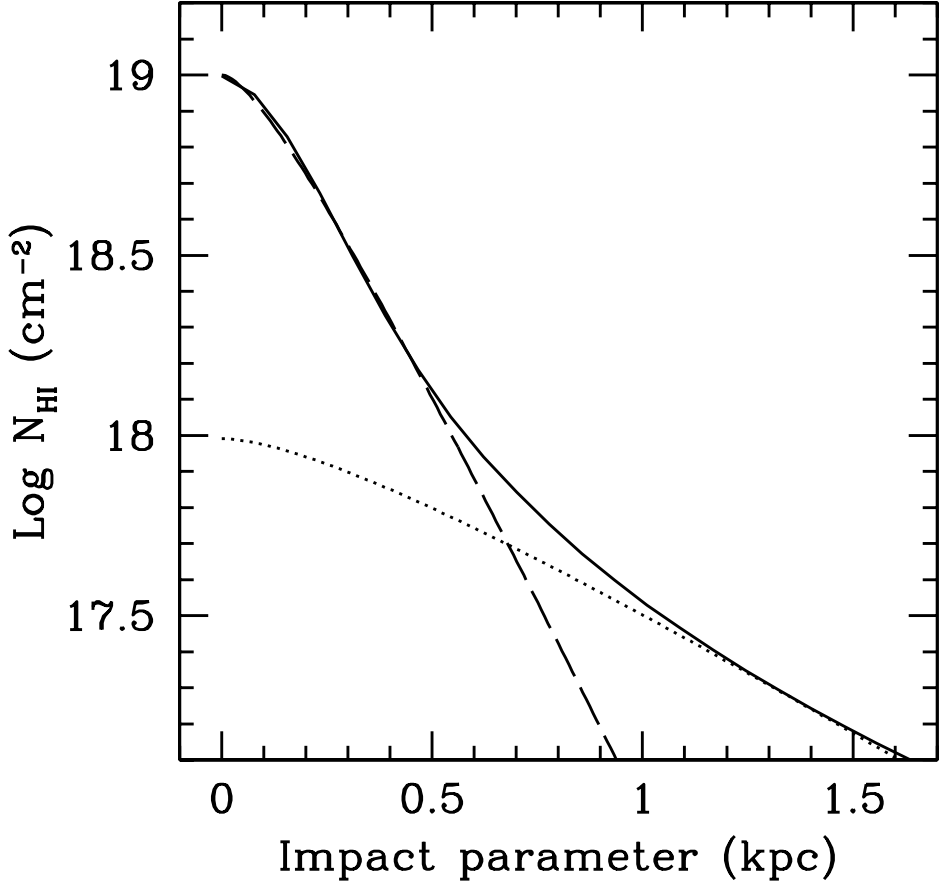,width=\textwidth,angle=0}}
\caption{The neutral hydrogen column density (solid lines) as a
function of impact parameter $b$ for a CHVC model in which the gas is
confined by a dark matter halo with a NFW profile. This halo has a
mass $\Mdm=5\times 10^8\msol$, and is shown as the solid line in the
center panel of Figure 7. Also plotted are fits for projected
exponential volume density distributions of neutral hydrogen, as
discussed in the text (dashed and dotted lines). The \Nhi\
distribution cannot be fit by a single profile as given by equation
(7) or (8). The inner and outer regions of the column density profile
are fit by infinite gas distributions (equation [7]) with $h=0.17$ kpc
and central column density $\Nhi(0)=1.0\times 10^{19}\psqcm$ (dashed
line) and $h=0.55$ kpc, $\Nhi(0)=9.8\times 10^{17}\psqcm$ (dotted
line), respectively.}
\end{figure}

\clearpage
\begin{figure}
\centerline{\psfig{figure=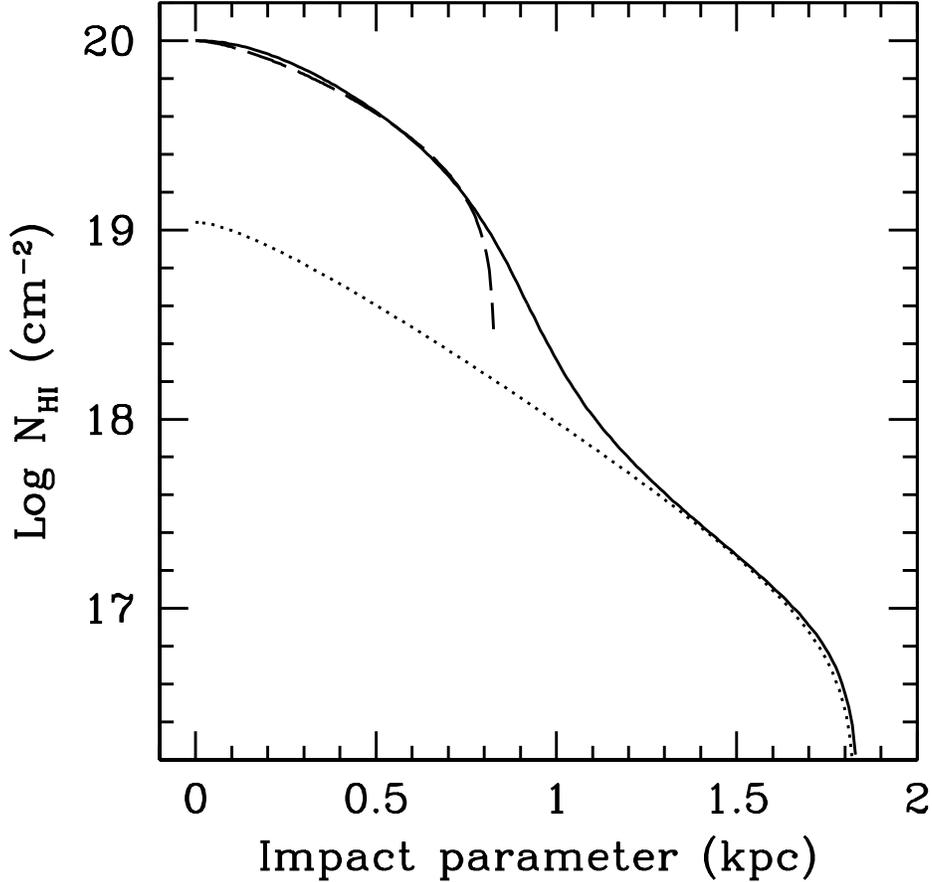,width=\textwidth,angle=0}}
\caption{The neutral hydrogen column density (solid lines) as a
function of impact parameter $b$ for a CHVC model in which the gas is
confined by a dark matter halo with a Burkert density profile. This
halo has a mass $\Mdm= 10^8\msol$, and is shown as the rightmost solid
line in the lefthand panel of Figure 8. Also plotted are fits for
projected exponential volume density distributions of neutral
hydrogen, as discussed in the text (dashed and dotted lines). The
\Nhi\ distribution cannot be fit by a single profile as given by
equation (7) or (8). The inner and outer regions of the column density
profile are fit by finite gas distributions (equation [8]) with
$h=0.4$ kpc and central column density $\Nhi(0)=10^{20}\psqcm$ (dashed
line) and $h=0.3$ kpc, $\Nhi(0)=1.1\times 10^{19}\psqcm$ (dotted
line), respectively. The cutoff radii for the two components are 0.83
kpc and 1.83 kpc, respectively.}
\end{figure}

\clearpage
\begin{table}
\caption[]{Results of CHVC models for an ionizing
  photon flux $\phi_i\sim 10^4$ \phoflux.\tablenotemark{a} 
\tablenotetext{a}{The columns give the volume density of total
  hydrogen (\nh), the total hydrogen column (\Nh) needed to produce a
  neutral column density of $10^{19}$, the cloud radius (\rh), the
  radius at the HIPASS sensitivity limit (\rhi), the mean neutral
  hydrogen volume density (\nhi), the distance (D) required to match
  the typical HIPASS angular size, the apparent mass that would be
  derived from HI observations (M$_{\rm HI}$), the total gas mass
  ($M_{gas}$), and the ratio of dark matter to gas mass needed to
  produce the typical observed line width of 25 \kms\ for {\it
  gravitational} confinement of the gas. (No entries are given for the
  $\nh \le 10^{-3}\pcubcm$ models, as they violate the line width
  constraint with no dark matter.)}}
\smallskip
\begin{tabular}{ccrrccccc}
\tableline
\tableline

\multicolumn{1}{c}{                                      $n_{H}$} &
\multicolumn{1}{c}{                                      $N_{H}$} &
\multicolumn{1}{c}{                                      $r_{H}$} &
\multicolumn{1}{c}{                                     $r_{\rm HI}$} &
\multicolumn{1}{c}{                                     $n_{\rm HI}$} &
\multicolumn{1}{c}{                                     D} &
\multicolumn{1}{c}{                                    $M_{\rm HI}$} &
\multicolumn{1}{c}{                                   $M_{gas}$} &
\multicolumn{1}{c}{                      $M_{dm}/M_{gas}$}   \\
\multicolumn{1}{c}{         $\scriptstyle (10^{-3}\; \rm cm^{-3})$} &
\multicolumn{1}{c}{         $\scriptstyle (10^{19}\; \rm cm^{-2})$} &
\multicolumn{1}{c}{                    $\scriptstyle (\rm kpc)$} &
\multicolumn{1}{c}{                    $\scriptstyle (\rm kpc)$} &
\multicolumn{1}{c}{              $\scriptstyle (10^{-3}\; \rm cm^{-3})$} &
\multicolumn{1}{c}{                            $\scriptstyle (kpc)$} &
\multicolumn{1}{c}{               $\scriptstyle (10^6\; \rm M_{odot})$} &
\multicolumn{1}{c}{               $\scriptstyle (10^6\; \rm M_{odot})$} &
\multicolumn{1}{c}{                          } \\

\tableline
 0.25 &  81.61 &   524  &  278  &  0.0058 & 47000 & 17000 & 4800000 & -- \\
 0.50 &  40.60 &   130  &   71  &  0.023 &  12000  &  1100 & 150000 & -- \\
 0.75 & 27.17 &   58   &   32   &  0.051 &  5400  &  220  &  20000 & -- \\
 1.0 &  20.50 &   33   &   18.7 &  0.087 &  3200  &  76  &  4800 & -- \\
 2.5 &  8.829 &   5.7  &   3.6  &  0.45 &  610  &  2.8  &  62	 & 4 \\
 5.0 &  5.074 &   1.6  &   1.14 &  1.4 &  190  &  0.28  &  2.8	 & 25 \\
 7.5 &  3.834 &   0.82 &   0.61 &  2.7 &  100  &  0.082  &  0.56 & 64 \\
10.0 &  3.227 &   0.52 &   0.40 &  4.1 &  67  &  0.035  &  0.19	 & 120 \\

\tableline
\\
\end{tabular}
\end{table}

\end{document}